\documentclass[a4paper,11pt]{article}
\usepackage{amssymb}
\usepackage{wasysym}
\usepackage{lscape}
\usepackage{pdflscape}
\usepackage{tabularx}
\newcommand{\ac}{\'}
\usepackage{epsfig,multicol,bbm}
\newcommand\fverb{\setbox\fverbbox=\hbox\bgroup\verb}
\newcommand\fverbdo{\egroup\medskip\noindent%
			\fbox{\unhbox\fverbbox}\ }
\newcommand\fverbit{\egroup\item[\fbox{\unhbox\fverbbox}]}
\newbox\fverbbox
\begin{document} 
\title{
How large is the contribution of cosmic web to $\Omega_\Lambda$ ? 
A preliminary study on a novel inhomogenous model}
\author{Stefano Viaggiu
\\
Dipartimento di Matematica, Universit\'a "Tor Vergata" Roma,\\
Via della Ricerca Scientifica, 1\\
Rome, Italy 00133,\\
viaggiu@axp.mat.uniroma2.it\\
Marco Montuori
\\
SMC-ISC-CNR and Dipartimento di Fisica,\\
Universit\'a "La Sapienza" Roma,\\
Ple. Aldo Moro 2\\
00185, Rome, Italy,\\
marco.montuori@roma1.infn.it,\\
}
\date{\today}\maketitle
\begin{abstract}
The distribution of matter in the universe shows a complex pattern, 
formed by cluster of galaxies, voids and filaments denoted as cosmic web.
Different approaches have been proposed to model such structure in the framework
of the general relativity. Recently, one of us has proposed a generalization ($\Lambda$FB model)
of the Fractal Bubble model, proposed by Wiltshire, 
which accounts for such large scale structure. The $\Lambda$FB model is an evolution of 
FB model and includes in a consistent way a description of inhomogeneous matter distribution and a $\Lambda$ term.
Here we analyze the $\Lambda$FB model focusing on the relation between cosmological parameters.
The main result is the consistency of $\Lambda$CDM model values for $\Omega_{\Lambda0}$ ($\approx 0.7$) and $\Omega_{k0}$ ($|\Omega_{k0}|<\approx 0.01$)
with a large fraction of voids. 
This allows to quantify to which extent the inhomogeneous structure could account for $\Lambda$ constant consistently with standard values of the other cosmological parameters. 
\end{abstract}

\section{Introduction}
The model of universe which currently gives the best fit of the available astrophysical observations is the $\Lambda$ Cold Dark Matter model ($\Lambda$CDM). The model is based on a exact solution of equation of General Relativity which assumes homogeneity and isotropy with a F-L metric. Quite soon, however, observational evidences required the addition of two main artefacts to the theory: the dark matter and the dark energy. Rotation curve of spiral galaxies, velocity 
dispersion of galaxies in galaxy clusters, cluster mass estimate from hot intra-cluster gas emission, gravitational lensing from galaxy cluster, large scale structure formation from CMB tiny fluctuations required an addition of dark matter. The observation of an accelerated expansion of the universe required the addition of the dark energy.
According to the best fit model, dark matter should account for $\approx 23\%$ of the total energy density, while the dark energy for $\approx 73\%$ (\cite{1,q,2}).
Notwithstanding, the nature of both remains unknown and is amongst the deepest problems of modern physics. For these reason in the past decade, several authors \cite{16}-\cite{C2}
have explored alternatives to the $\Lambda$CDM model.\\
The discovery of a lumpier universe than expected offered a possibility in this respect. The universe appears indeed organized as a cosmic web, which can be described as an interconnected network: spherical clusters form the nodes and are joined by elongated filaments defining 2D sheets. Recent analysis have shown that voids in the network could fill between $\approx 40\% -70\%$ of the space (accordingly to definition and measure of the voids) and have a continuous distribution of sizes depending on the galaxy sample selected \cite{web, 100,100v,101v}. Such observed inhomogeneities are usually described as a first order perturbation with respect to the homogeneous and isotropic exact solution. This formulation has a limited application, since it is valid for small perturbation $\delta \rho/\rho$ and cannot describe the lumpy universe at low redshift. Nevertheless, it was soon clear, mainly through the formulation of the Buchert formalism, that such an inhomogeneous matter distribution could indeed mimic the presence of dark energy \cite{102,103,104,u1,u2}. Since then, some authors explored this possibility to the extreme consequence to explain the whole amount of dark energy on the basis of inhomogeneous distribution. Our personal opinion is that the requested existence of Gpc scale voids makes these models questionable. From the other side, the standard perturbative approach of $\Lambda$CDM cannot answer to the question of the effect of small scale strong fluctuations on larger scale.
In this context, an interesting line appears the exploration of cosmological models including the largely accepted
cosmological constant paradigm and at the same time a non perturbative approach to the observed inhomogeneities.
This issue is a very recent one (see \cite{aer}) and so far limited to the not so realistic (even useful) case of spherical simmetry.
Going beyond such studies and toward a more realistic model of matter distribution in the universe is the target of the present paper.
The elaboration of a similar model will allow, for example, to study in a non perturbative way the percentage of dark energy that can be explained in terms of the observed inhomogeneities or the role of voids in the process of structure formation (see for example \cite{void}) in a more complete way for the presence both of dark energy and voids in the model.
More in general, it would be possible to explore the effect of the observed inomogeneities onto cosmic scale data, as barionic acoustic oscillations and cosmic radiation.

Our baseline is the so-called 'fractal bubble'
FB or 'timescape' cosmology \cite{10,12}. It is based on the Buchert average scheme \cite{5} and introduces a non-uniform time flow
with two scales corresponding to the voids and walls of a schematic lattice cosmic structure.
In the FB model the apparent acceleration is explained in terms
of different rate of clocks located in walls (our point of observation) as compared to the clocks of voids.
The  viability of the FB or Timescape model as an alternative to the $\Lambda$CDM model is still matter of debat \cite{ww1,ww2}.
In any case, the FB model is the first one which describes a schematic cosmic web in a non perturbative way and without
invoking particular symmetries. This interesting feature has prompted one of us to propose an inhomogeneous model based on the same partitioning within the Buchert average scheme,
but including a $\Lambda\neq 0$ and an uniform time flow \cite{sv}. This model, hereafter $\Lambda$FB model, allows to investigate,  
in a non perturbative way,
the effects of large scale spatial inhomogeneities and non-vanishing curvature on the cosmological parameters. 
The present preliminary study is devoted to such aim. The scheme of the paper is the following: in section 2 we write
the relevant equations of the model. We obtain a new interesting equation involving all the parameters of the model
evaluated at the present epoch, which is analysed in sections 3 and 4.
In section 5 we 'dress' the cosmological parameters and finally, we report our conclusions in Section 6.

\section{Basic equations}
The $\Lambda$FB model \cite{sv} incorporates the observed present inhomogeneities within the cosmological constant
paradigm. The starting point is the FB model introduced by Wiltshire \cite{10,12}. This model considers a matter distribution in the universe 
as a regular network formed by walls and voids. 
The $\Lambda$FB model assumes the same partitioning of the two scale FB model
\cite{12}: regions called 'finite infinity' ($F_I$) which are the
set of timelike boundaries of compact
{\bf disjoint} domains
$I$, with a zero average expansion and a positive one outside,
i.e. ${<\theta>}_{{\bigcup}_{I}F_I}=0$. The $F_I$ regions are within 'wall regions'
whose metric is, on average:
\begin{equation}
ds^2_w=-dt^2+a^2_w\left[d{\eta}^2_w+{\eta}^2_wd{\Omega}^2\right].
\label{1}
\end{equation}
Our position is in $F_I$. The regions complementary to
the walls with respect to the particle horizon are called voids and are expanding with an average hyperbolic metric given by:
\begin{equation}
ds^2_v=-dt^2+a^2_v\left[d{\eta}^2_v+{\sinh}^2({\eta}_v)d{\Omega}^2\right].
\label{2}
\end{equation}
Note that, contrary to the FB model, in our model the time flow is isotropic. 
This is due to the presence of a non vanishing cosmological constant, which is absent in the FB model 
since it intends to explain it as an effect of inhomogeneity.
The isotropic flow has the useful feature to avoid the shortcomings
related to the junction conditions of the original FB model (see \cite{sv2}).
In any case, a nontrivial phenomenological
lapse function involves new physics relating to gravitational energy which
is not yet widely accepted, and we will therefore consider the commonly
accepted alternative that time flows uniformly.\\
The Hubble parameters in walls and voids are respectively
$H_w=\frac{{<\theta>}_w}{3}=\frac{{\dot{a}}_w}{a_w},\;
H_v=\frac{{<\theta>}_v}{3}=\frac{{\dot{a}}_v}{a_v}$.
An important assumption of the $\Lambda$FB model (and FB one) is the existence of a scale of
homogeneity with the scale factor $a(t)$. 
The average on the whole volume of the particle horizon $V=a^3 V_i$ is:
\begin{equation}
a^3=f_{wi}\;a_{w}^3+f_{vi}\;a_{v}^3,\;\;\;f_{wi}+f_{vi}=1,
\label{3}
\end{equation}
where $f_{wi}$ and $f_{vi}$ are the fractions of walls and voids at the time $t=t_i$. 
We choose the time $t_i$ (the initial time) as the recombination time in order to compare the model with available observational data.
In general the following relations are valid for any time:
\begin{equation}
f_v(t)+f_w(t)=1,\;\;f_w=f_{wi}\frac{a_{w}^3}{a^3},\;\;f_v=f_{vi}\frac{a_{v}^3}{a^3}.
\label{4}
\end{equation}
The Hubble rate $H$ at the scale of homogeneity is:
\begin{equation}
H=f_w H_w+f_v H_v,\;\;H=I_w H_w=I_v H_v.
\label{5}
\end{equation}
and the density parameters are:
\begin{equation}
{\Omega}_m=\frac{8\pi<\rho>}{3H^2},\;{\Omega}_k=-\frac{<\mathcal{R}>}{6H^2},\;
{\Omega}_{\mathcal{Q}}=-\frac{\mathcal{Q}}{6H^2},\;
{\Omega}_{\Lambda}=\frac{\Lambda}{3H^2},
\label{6}
\end{equation}
where $\mathcal{Q}$ is the kinematic backreaction and $\mathcal{R}$ is the
spatial curvature.
The Buchert equations are:
\begin{equation}
{\Omega}_m+{\Omega}_{\Lambda}+{\Omega}_k+{\Omega}_{\mathcal{Q}}=1,\;
\left(6\mathcal{Q}+2<\mathcal{R}>\right)\dot{a}+a\left[\dot{\mathcal{Q}}+
<\dot{\mathcal{R}}>\right]=0.
\label{7}
\end{equation}
Equation (\ref{7}) can be easily manipulated to obtain the final equations for the model (for more details see
\cite{sv}):
\begin{eqnarray}
& &(1-f_v)\frac{\dot{a}}{a}-\frac{{\dot{f}}_v}{3}=
\sqrt{{\Omega}_{m0}H_0^2\frac{a_0^3}{a^3}\left(\frac{1-{\epsilon}_{i}}{{\Omega}_F}
\right)(1-f_v)},\label{8}\\
& &\frac{\dot{a}}{a}+\frac{{\dot{f}}_v}{3f_v}=\frac{a_0 H_0}{a{f}_v^{\frac{1}{3}}}
\sqrt{\frac{{\Omega}_{k0}}{f_{v0}^{\frac{1}{3}}}+
\frac{{\Omega}_{\Lambda0}a^2}{f_v^{\frac{1}{3}}a_0^2}
+\frac{a_0{\Omega}_{m0}}{a{f}_v^{\frac{1}{3}}}
\left(1+\frac{{\epsilon}_{i}-1}{{\Omega}_F}\right)}, \label{9}
\end{eqnarray}
where ${\epsilon}_i$ is a integration constant.
A first integral of eqs (\ref{8}, \ref{9}) is:
\begin{equation}
\frac{(1-{\epsilon}_{i})I_w^2{\Omega}_m}{(1-f_v)}=
{\Omega}_F=e^{-\int_{t_i}^{t}\frac{\Lambda}{H_w}dt}.
\label{10}
\end{equation}
From equation (\ref{8}) (see \cite{sv})
we can solve for $a_w$ and ${\Omega}_F$:
\begin{equation}
a_w= a_{w0}\;{\sinh}^{\frac{2}{3}}\left(\frac{\sqrt{3\Lambda}}{2}t\right),\;
{\Omega}_F={\left(\frac{\cosh\left(\sqrt{\frac{3\Lambda}{4}}\;t_i\right)}
{\cosh\left(\sqrt{\frac{3\Lambda}{4}}\;t\right)}\right)}^2.
\label{11}
\end{equation}
We are now able to derive an useful and interesting formula.
By combining equations (\ref{8}) and (\ref{9}) and evaluating them at the present epoch $t_0$
we obtain:
\begin{equation}
1=\sqrt{{\Omega}_{m0}(1-f_{v0})\frac{(1-{\epsilon}_i)}{{\Omega}_{F0}}}
+\sqrt{f_{v0}}\sqrt{{\Omega}_{k0}+{\Omega}_{\Lambda0}+
{\Omega}_{m0}\left(1+\frac{{\epsilon}_i-1}{{\Omega}_{F0}}\right)},
\label{12}
\end{equation}
The equation (\ref{12}) constraints the density parameters and the fraction of voids at the present time $t_0$. 
The equation was already present in \cite{sv}. Now we intend to go a step beyond, expressing the constant 
${\Omega}_{F0}$ in terms of density parameters.
Integrating equation
(\ref{8}) we get:
\begin{equation}
{(1-f_v)}^{\frac{1}{3}} a=
a_0{\left(\frac{{\Omega}_{m0}(1-{\epsilon}_i)}{{\Omega}_{\Lambda0}}\right)}^{\frac{1}{3}}
\frac{{\sinh}^{\frac{2}{3}}\left(\frac{\sqrt{3\Lambda}}{2}t\right)}
{{\cosh}^{\frac{2}{3}}\left(\frac{\sqrt{3\Lambda}}{2}t_i\right)}.
\label{13}
\end{equation}
It is should be noticed that in the equation (\ref{13}), according to \cite{10} and without loss
of generality, we have set to zero the integration constant. 
In such a way, the study of the cosmological constant is simpler. This choice constraints the integration constant in the integration of the equation (\ref{9}) (see \cite{10}).
Moreover, since the current estimation of $\Lambda$ is very small and $t_i \approx 3.77\cdot 10^{-5}$ Gyr, we can set with good approximation ${{\cosh}^{\frac{2}{3}}\left(\frac{\sqrt{3\Lambda}}{2}t_i\right)}\simeq 1$.
Evaluating equation (\ref{13}) at the present epoch, we get:
\begin{equation}
\frac{(1-{\epsilon}_i)}{{\Omega}_{F0}}=
1-{\epsilon}_i+\frac{{\Omega}_{\Lambda0}}{{\Omega}_{m0}}\left(1-f_{v0}\right).
\label{14}
\end{equation}
and by putting (\ref{14}) in (\ref{12}) we have:
\begin{eqnarray}
& &1=\sqrt{(1-f_{v0})[{\Omega}_{m0}(1-{\epsilon}_i)+{\Omega}_{\Lambda0}(1-f_{v0})]}+\nonumber\\
& &+\sqrt{f_{v0}}\sqrt{{\Omega}_{k0}+{\Omega}_{m0}{\epsilon}_i+f_{v0}{\Omega}_{\Lambda0}}.
\label{15}
\end{eqnarray}
The formula (\ref{15}) was not present in \cite{sv}; it involves all the parameters of the $\Lambda$FB model and has been obtained by
using all the relevant equations (i.e. (\ref{8}) and (\ref{9})). Moreover, it gives a compatibility
equation for cosmological parameters that are averaged quantities in principle measurable. It should be noted that
this equation is not perturbative. It allows to study the effects of the inhomogeneities
without ruling out the dark energy and avoiding the shortcomings of the perturbation theory when applied to large inhomogeneities. \\
The cosmological parameters present in (\ref{15}) are different from those of the $\Lambda$CDM model. The concordance model is based on an exact solution of Einstein's equations and the corresponding cosmological parameters are
local quantities. The homogeneity and isotropy of the solution implies that the cosmological parameters are the same in any spatial point for any fixed comoving time $t$. As the timescape model, also the $\Lambda$FB one is obtained within the Buchert 
formalism, where the cosmological parameters are averaged non local quantities. In this formalism for a given scalar $\psi$ at a fixed cosmological time $t$, the average on the whole particle 
horizon is:
\begin{equation}
<\psi(t)>=f_w(t){\psi}_w(t)+f_v(t){\psi}_v(t),
\label{mean}
\end{equation}
where the subscrits $v$ and $w$ refer obviously to the values of  $\psi$ in voids and walls.

Astrophysical data refer  to the local value of $\psi$ that can be generally different from its averaged one $<\psi>$.
If we consider the universe as a cosmic web, these averaged parameters are physical quantities
obtained from an average procedure over a given astrophysical sample containing voids and walls.
Moreover, ${\psi}_w(t)$ and   ${\psi}_v(t)$ are mean values 
calculated separately in walls and voids.\\
In any case, the model has a scale of statistical homogeneity, beyond which it has a Friedmann metric evolving with $a(t)$ given by eq.(\ref{3}).
By consequence, the cosmological parameters obtained as averaged quantities at the scale 
of homogeneity can be straightforward compared with the corresponding values of the standard concordance model. 
For this reason, we expect that the numerical
values for the cosmological parameters of the ${\Lambda}$FB model are similar to those of the standard cosmological model.\\
As a further consideration on the $\Lambda$FB model, note that
the formalism introduced allows easily to compute the limit for ${\Omega}_{k0}\rightarrow 0$ in (\ref{8})-(\ref{9}). At this aim 
we should just change the hyperbolic metric (\ref{2}) with parabolic void metric.
\begin{equation}
ds^2_v=-dt^2+a^2_v\left[d{\eta}^2_v+{\eta}^2_vd{\Omega}^2\right].
\label{15p}
\end{equation}
This case is an interesting one: if ${\Omega}_{k0}=0$ and ${\Omega}_{m0}+{\Omega}_{\Lambda0}>1$, 
equation (\ref{7}) implies that ${\Omega}_{\mathcal{Q}}\leq 0$. The latter means that the flat void region described by metric (\ref{15p}) 
should be an underdensity. Equation (\ref{15p}) is relevant only to compute the distance-redshift relation, i.e. when considering data on the  light cone. Then, 
we can study equation (\ref{15}) also  in the limit
${\Omega}_{k0}=0$. In this case, we get the $\Lambda$CDM model by setting $f_{v0}={\epsilon}_i=0$; in another words, at odds with FB model, the $\Lambda$FB contains the $\Lambda$CDM model for the aforementioned choice of the parameters.

\section{Constraints on the parameter ${\epsilon}_i$}
From equation (\ref{5}) and (\ref{10}) we have (for $t=t_i$):
\begin{equation}
{\epsilon}_i=1-\frac{(1-f_{vi})}{I_{wi}^2{\Omega}_{mi}},\;\;
I_{wi}=1-f_{vi}+f_{vi}\frac{H_{vi}}{H_{wi}}=\frac{H_i}{H_{wi}}.
\label{15q}
\end{equation}
In the original FB model, the parameter $I_w$ is interpreted as a phenomenological lapse function and we have 
$I_{wi}= 1, {\Omega}_{mi}\simeq 1,{\epsilon}_i<<1$. By consequence, in the FB model the parameter
${\epsilon}_i$ has no role. On the contrary, in our model ($\Lambda$FB), $I_w$ is not a lapse function but simply a measure
of the ratio $\frac{H}{H_w}$ or in another words of the ratio between the expansion rate of voids and walls. By consequence, ${\epsilon}_i$ is only constrained to be $\leq 1$, from the existence of II member of eq.(\ref{8}). It is also possible to express,
by means of (\ref{15}), ${\epsilon}_i$ in terms of the other current cosmological parameters. We have
\begin{eqnarray}
& &{\epsilon}_i=\frac{-b\pm 4{\Omega}_{m0}\sqrt{\Delta}}{2a} \label{16q}\\
& &b=-2{\Omega}_{m0}{\Omega}_{\Lambda0}+2f_{v0}{\Omega}_{m0}{\Omega}_{k0}+
4f_{v0}{\Omega}_{m0}{\Omega}_{\Lambda0}-4f_{v0}{\Omega}_{m0}-\nonumber\\
& &-2{\Omega}_{m0}^2+2f_{v0}{\Omega}_{m0}^2+2{\Omega}_{m0},\nonumber\\
& &a={\Omega}_{m0}^2,\;\;\
\Delta=f_{v0}({\Omega}_{k0}+{\Omega}_{m0}+{\Omega}_{\Lambda0}-1)(1-f_{v0}).\nonumber
\end{eqnarray}
Care must be taken to avoid spurious solutions. In particular, for $f_{v0}\leq 0.1$
we generally have one root, the greater one in (\ref{16q}).
The existence of the solution requires ${\Omega}_{m0}+{\Omega}_{\Lambda0}+{\Omega}_{k0}\geq 1$, which implies for 
equation \ref{7} ${\Omega}_{\mathcal{Q}}\leq 0$, i.e.
a non positive backreaction. 
This is a check of the consistency of the model, since the partition chosen with voids with 
an average negative curvature should imply a non positive backreaction, as it is.
\section{Constraints on the cosmological parameters of the $\Lambda$FB model}
\subsection{The current curvature ${\Omega}_{k0}$}
In this section we discuss the constraints on the cosmological parameters we can get from the observational data.
Equation (\ref{15}) can be solved with respect to ${\Omega}_{k0}$ and we get:
\begin{equation}
{\Omega}_{k0}=\frac{{\left[1-\sqrt{(1-f_{v0})[{\Omega}_{m0}(1-{\epsilon}_i)
+{\Omega}_{\Lambda0}(1-f_{v0})]}\right]}^2}{f_{v0}}-
f_{v0}{\Omega}_{\Lambda0}-{\Omega}_{m0}{\epsilon}_i.\label{165q}
\end{equation}
Current estimates of ${\Omega}_{k0}$ come from CMB and SZ effect data (see \cite{106,107,190}). Such measures can be interpreted 
in the standard framework ($\Lambda$CDM) or in the void models (as the FB, the present $\Lambda$FB, etc..).
In the $\Lambda$CDM model, which assumes a constant spatial curvature at fixed time, the observations imply $|{\Omega}_{k0}|<0.01$.
Remember that within the Buchert scheme, this limit can also be taken into account, provided that it is intended as a mean quantity.
 Such small value is also consistent with the inflationary scenario. On the contrary in void models, the curvature acts as an effective dark energy. For this reason and in order to fit the observational data, void models require a quite large curvature. Such large value has been claimed consistent to the observation of big voids in the large scale structure. Unfortunately such scenario has two additional implications: a low value for the Hubble constant and/or the existence of a giant void around our location 
(see \cite{106T}). 
Both of them appear quite implausible. The present model $\Lambda$FB on the contrary, since incorporates voids and dark energy, gives the possibility to explore the compatibility between dark energy, large voids, a standard range of values for ${\Omega}_{m0}\in [0.25,0.35]$ and a small value for ${\Omega}_{k0}$
in agreement with the WMAP constraint.
From the formula (\ref{16q}) we see that a relevant volume void fraction $f_{v0}$ is compatible with a rather small value of ${\Omega}_{k0}$. 
For examples considering ${\Omega}_{\Lambda0}=0.7$ and ${\Omega}_{m0}=0.3$ we have:
\begin{itemize}
\item $f_{v0}=0.7$ for ${\Omega}_{k0}=0.0039,{\epsilon}_i=0.5$
\item $f_{v0}=0.5$ for ${\Omega}_{k0}=0.001, {\epsilon}_i=0.16$
\item $f_{v0}=0.2$ for ${\Omega}_{k0}=0.01, {\epsilon}_i=0.46, {\epsilon}_i=-0.07$
\end{itemize}
Note that can exist values for $f_{v0}$ obtained with two values for ${\epsilon}_i$.\\
Decreasing ${\Omega}_{\Lambda0}$ in order to get a similar value for $f_{v0}$, it
requires a larger ${\Omega}_{k0}$, in agreement with the inhomogeneous models where dark energy is mimicked by the curvature.\\
Note that for the same range of parameter values, the backreaction term is ${\Omega}_{\mathcal{Q}} \approx -0.06$ and thus remains relatively small.\\
Larger values for ${\Omega}_{\mathcal{Q}}$ can be obtained  with a small ${\Omega}_{k0}$ and a reasonable void fraction $f_{v0}$, for a smaller ${\Omega}_{\Lambda0}$, but a larger ${\Omega}_{m0}$ value.
It is worth to stress that our result is not a validation of the standard value for ${\Omega}_{k0}$.
In fact, the current standard constraint $|{\Omega}_{k0}|\leq 0.01$ is obtained within the Friedmann paradigm. 
A larger value ${\Omega}_{k0} \approx 1$ can fit as well the experimental data (CMB and BAO), but it is at odds with inflationary paradigm. 
Our result is indeed a falsification of a common claim, i.e. large density contrast
$|{\delta\rho}/\rho| \simeq 1$ due to voids necessarily implies a large curvature contrast
$|{\delta \mathcal{R}}/\mathcal{R}|\simeq 1$. In our model a large fraction of voids is compatible with a small spatial 
curvature and a standard value for cosmological constant.

\subsection{Current volume voids fraction $f_{v0}$}
An interesting feature of the $\Lambda$FB model is the constraints on the present day volume void fraction $f_{v0}$ from the values of the other cosmological parameters .
By posing the condition ${\Omega}_{k0}\geq 0$, ${\Omega}_{m0}>-2{\Omega}_{\Lambda0}+2\sqrt{{\Omega}_{\Lambda0}}$ (which is always satisfied)
and neglecting spurious solutions we have
\begin{eqnarray}
& &f_{v0}\geq X_0,\;\;\;X_0=\frac{-b+4\sqrt{\Delta}}{2a},\label{17q}\\
& &a=4{\Omega}_{\Lambda0}^2-4{\Omega}_{\Lambda0}+{\Omega}_{m0}^2+
4{\Omega}_{m0}{\Omega}_{\Lambda0},\nonumber\\
& &\Delta=({\Omega}_{m0}+{\Omega}_{\Lambda0}-1)
\left[{\Omega}_{\Lambda0}({\Omega}_{m0}+{\Omega}_{\Lambda0}-1)+ 
{\epsilon}_i{\Omega}_{m0}^2(1-{\epsilon}_{i})\right],\nonumber\\
& &b=-4{\Omega}_{m0}{\epsilon}_i+4{\Omega}_{m0}{\Omega}_{\Lambda0}{\epsilon}_{i}+2{\Omega}_{m0}^2{\epsilon}_{i}-
2{\Omega}_{m0}^2-6{\Omega}_{m0}{\Omega}_{\Lambda0}+
4{\Omega}_{\Lambda0}+\nonumber\\
& &+2{\Omega}_{m0}-4{\Omega}_{\Lambda0}^2.\nonumber
\end{eqnarray}
Note that, if ${\Omega}_{m0}+{\Omega}_{\Lambda0}<1$, no limitations arise for the actual volume void fraction $f_{v0}$.
For ${\Omega}_{m0}+{\Omega}_{\Lambda0}=1$ we obtain $X_0={\epsilon}_{i}$.\\
Finally, note that in the case ${\Omega}_{k0}=0$ (for all times), the inequality (\ref{17q})
becomes an equality. As a result, in this case we can obtain the standard concordance model ($f_{v0}=0$)
or a model with voids with a flat Friedmann metrics and $f_{v0}=X_0$, provided that
${\Omega}_{m0}+{\Omega}_{\Lambda0} >1$.\\
We study now the equation (\ref{15}) in terms of the allowed current volume voids fraction $f_{v0}$.
The equation can be solved in terms of $f_{v0}$ with some little algebra.
Care must be taken for the possible appearance of spurious
solutions, by solving it graphically (i.e. equation (\ref{15})) and eliminating the latter ones.
Then expliciting equation (\ref{15}) in terms of $f_{v0}$ we get:
\begin{eqnarray}
& &f_{v0}=\frac{-b\pm 4\sqrt{\Delta}}{2a},\label{vuoti}\\
& &a=4{{\Omega}}_{\Lambda0}^2+{{\Omega}}_{m0}^2+{{\Omega}}_{k0}^2-4{\Omega}_{\Lambda0}+
4{\Omega}_{\Lambda0}{\Omega}_{k0}+2{\Omega}_{m0}{\Omega}_{k0}+
4{\Omega}_{m0}{\Omega}_{\Lambda0},\nonumber\\
& &b=4{\Omega}_{\Lambda0}+2{\Omega}_{m0}-6{\Omega}_{\Lambda0}{\Omega}_{m0}-2{\Omega}_{k0}-
2{\Omega}_{\Lambda0}{\Omega}_{k0}-2{{\Omega}}_{m0}^2-4{{\Omega}}_{\Lambda0}^2-\nonumber\\
& &-2{\Omega}_{k0}{\Omega}_{m0}
+4{\Omega}_{m0}{\Omega}_{\Lambda0}{\epsilon}_{i}+2{\Omega}_{m0}^2{\epsilon}_{i}+
2{\Omega}_{k0}{\Omega}_{m0}{\epsilon}_{i}-4{\Omega}_{m0}{\epsilon}_{i},\nonumber\\
& &\Delta=({\Omega}_{\Lambda0}+{\Omega}_{m0}+{\Omega}_{k0}-1)
[{\Omega}_{m0}{\Omega}_{\Lambda0}+{\Omega}_{\Lambda0}^2+
{\Omega}_{\Lambda0}{\Omega}_{k0}-{\Omega}_{\Lambda0}+\nonumber\\
& &+{\Omega}_{m0}^2{\epsilon}_{i}(1-{\epsilon}_{i})
+{\Omega}_{m0}{\Omega}_{k0}(1-{\epsilon}_{i})].\nonumber
\end{eqnarray}
Note that for ${\Omega}_{\Lambda0}+{\Omega}_{m0}>1$ at least a solution is always present, while for
${\Omega}_{\Lambda0}+{\Omega}_{m0}<1$ we can have no solutions. In any case for
${\Omega}_{m0}+{\Omega}_{\Lambda0}+{\Omega}_{k0}<1$, since ${\Omega}_{\mathcal{Q}}<0$, 
according to equation (\ref{7}), no solutions are available.
For an example of possible values, see the table (\ref{voids}).
\begin{table}[t]
\begin{center}
\begin{tabular}{|c|c|c|c|c|c|}
\hline
${\Omega}_{\Lambda0}$ & ${\Omega}_{m0}$ & ${\Omega}_{k0}$ & ${\epsilon}_{i}$ &
$f_{v_{01}}$ & $f_{v_{02}}$\\
\hline
$0.7$ & $0.28$ & $0.01$ & $0$  & $no$ & $no$\\
$0.7$ & $0.28$ & $0.02$ & $0$ & $no$ & $0.066$\\
$0.7$ & $0.28$ & $0.021$ & $0$ & $0.03$ & $0.14$\\
$0.7$ & $0.28$ & $0.022$ & $0$ & $0.023$ & $0.18$\\
$0.7$ & $0.29$ & $0.018$ & $0$ & $0.0026$ & $0.32$\\
$0.7$ & $0.29$ & $0.018$ & $0.01$ & $0.004$ & $0.33$\\
$0.7$ & $0.30$ & $0.001$ & $0$ & $no$ & $0.043$\\
$0.7$ & $0.30$ & $0.01$ & $0$ & $no$ & $0.32$\\
$0.7$ & $0.30$ & $0.02$ & $0$ & $no$ & $0.51$\\
$0.7$ & $0.30$ & $0.01$ & $0.01$ & $0.0002$ & $0.34$\\
$0.7$ & $0.30$ & $0.01$ & $0.1$ & $0.016$ & $0.44$\\
$0.7$ & $0.305$ & $0.0007$ & $0$ & $no$ & $0.18$\\
$0.7$ & $0.305$ & $0.01$ & $0$ & $no$ & $0.4$\\
$0.7$ & $0.31$ & $0.01$ & $0$ & $no$ & $0.47$\\
$0.7$ & $0.31$ & $0.01$ & $0.01$ & $no$ & $0.48$\\
$0.7$ & $0.31$ & $0.01$ & $0.15$ & $0.014$ & $0.62$\\
\hline
\end{tabular}
\caption{Allowed actual volume voids fraction for $\Lambda$FB model.}
\label{voids}
\end{center}
\end{table}
Quite generally, we can have two possible values for $f_{v0}$: 
\begin{itemize}
\item $f_{v01} <<1$ 
\item $f_{v02}\approx 0.1$ or greater
\end{itemize}
This implies that if the initial fraction of voids is:
\begin{itemize}
\item $f_{vi} < f_{v01}$ then at $t=t_0$ $f_{v0} =f_{v01} $
\item$f_{vi} > f_{v01}$ then at $t=t_0$ $f_{v0} =f_{v02} $
\end{itemize}
Considering the value for $f_{vi}$ estimated from WMAP data 
($\approx\in [10^{-5},10^{-2}]$) 
and the values of $f_{v01}$ and $f_{v02}$, both of the previous cases could be fullfilled. According to the model, starting from a fraction of voids $f_{vi}$ within WMAP constraints,
we could get at the present time $t_0$ both $f_{v01}$ and $f_{v02}$. Which one is a question of the precise value of $f_{vi}$ with respect to the $f_{v01}$. The dependence from the other parameters is quite complex. In fact from the table (\ref{voids}) is apparent that for a choice ${\Omega}_{\Lambda0}\simeq 0.7$, ${\Omega}_{k0}\leq 0.02$  and $ 0\leq  {\epsilon}_i\leq 0.15$, changing ${\Omega}_{m0}$ from $0.28$ to $0.31$ produces a variation of $f_{v0}$ from $\sim 0.0002$ up to $\simeq 0.6$. 
In the $\Lambda$FB, contrary to the
FB one, the final fraction of voids $f_{v0}$ is dependent from the initial one $f_{vi}$. This happens because in the  
 $\Lambda$FB model no tracker solutions are available.
\subsection{Current cosmological constant ${\Omega}_{\Lambda0}$}
\begin{figure}[h!]{\psfig{file=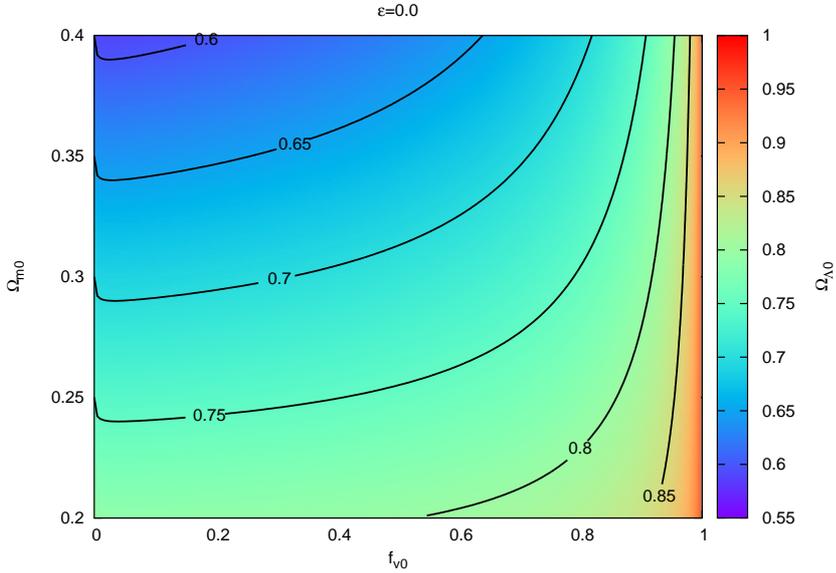,width=8cm, angle=-90} 
        \caption{${\Omega}_{\Lambda0}$ vs. ${\Omega}_{m0}$ and ${f}_{v_0}$. ${\epsilon}_i = 0$. The labelled curves are the isocurves for the corresponding value of ${\Omega}_{\Lambda0}$.}
	\label{fig1}}
\end{figure}

\begin{figure}[h!]{\psfig{file=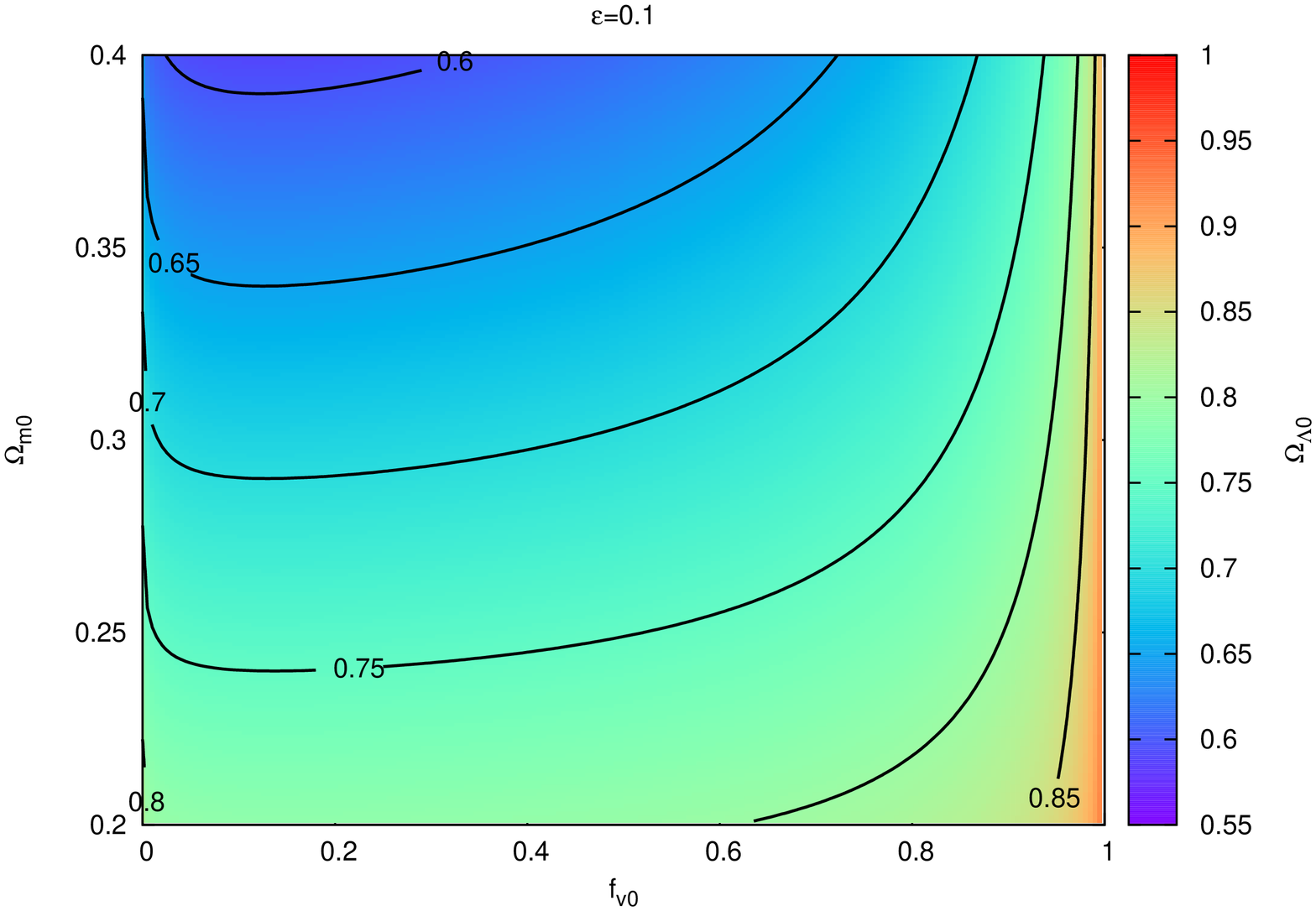,width=8cm, angle=-90} 
        \caption{${\Omega}_{\Lambda0}$ vs. ${\Omega}_{m0}$ and ${f}_{v_0}$. ${\epsilon}_i = 0.1$. For the isocurves see caption fig\ref{fig1} }
	\label{fig2}}
\end{figure}

\begin{figure}[h!]{\psfig{file=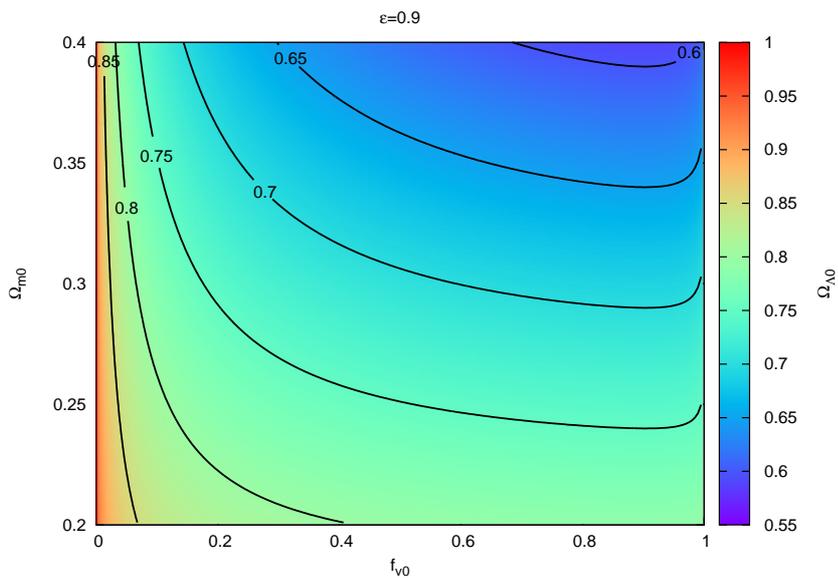,width=8cm, angle=-90} 
        \caption{${\Omega}_{\Lambda0}$ vs. ${\Omega}_{m0}$ and ${f}_{v_0}$. ${\epsilon}_i = 0.9$.For the isocurves see caption fig\ref{fig1} }
	\label{fig3}}
\end{figure}

As mentioned, both FB and $\Lambda$FB models collapsing regions
are within walls; for this reason the model has ${\Omega}_{k0} \geq 0$ 
and we get:
\begin{eqnarray}
& &{\Omega}_{\Lambda0} \leq G \label{18}\\
& &G=\frac{{\epsilon}_{i}{\Omega}_{m0}-
2f_{v0}{\epsilon}_{i}{\Omega}_{m0}+1-2{\Omega}_{m0}f_{v0}^2-2f_{v0}+3f_{v0}{\Omega}_{m0}-
{\Omega}_{m0}+2f_{v0}^2-2\sqrt{Q}}{{(1-2f_{v0})}^2}, \nonumber\\
& &Q=f_{v0}(1-f_{v0})[2f_{v0}^2{\Omega}_{m0}-f_{v0}^2+f_{v0}-
f_{v0}{\Omega}_{m0}+
{\epsilon}_{i}{\Omega}_{m0}(1-2f_{v0})].\nonumber
\end{eqnarray}
Note that $Q>0$ for all values allowed of $f_{v0}, {\Omega}_{m0}$.
We conclude this subsections by analysing
the formula (\ref{15}) from the point of view of ${\Omega}_{\Lambda0}$. This is in our opinion the main point of the present study, since it allows to estimate the fraction of ${\Omega}_{\Lambda0}$ due to the inhomogeneities in the large scale structure. 
By solving (\ref{15})
with respect to ${\Omega}_{\Lambda0}$ we obtain
\begin{eqnarray}
& &{\Omega}_{\Lambda0}= \frac{-b-4\sqrt{f_{v0}(1-f_{v0})}\sqrt{\Delta}}{2a}\label{L1},\\
& &a=1+4f_{v0}^2-4f_{v0},\nonumber\\
& &b=-6{\Omega}_{m0}{f}_{v_0}+4{f}_{v_0}-2+4{f}_{v_0}{\Omega}_{m0}{\epsilon}_{i}-
4{f}_{v_0}^2+2{\Omega}_{m0}+4{f}_{v_0}^2{\Omega}_{m0}+\nonumber\\
& &+4{f}_{v_0}^2{\Omega}_{k0}-
2{f}_{v_0}{\Omega}_{k0}-2{\Omega}_{m0}{\epsilon}_{i},\nonumber\\
& &\Delta=2f_{v0}^2{\Omega}_{m0}-f_{v0}^2+f_{v0}-
f_{v0}{\Omega}_{m0}+
{\epsilon}_{i}{\Omega}_{m0}(1-2f_{v0})+2{f}_{v_0}^2{\Omega}_{k0}-\nonumber\\
& &-3f_{v0}{\Omega}_{k0}+
{\Omega}_{k0}.\nonumber
\end{eqnarray}
To our knowledge this is the first equation which allows to compute the behaviour of ${\Omega}_{\Lambda0}$ versus 
in particular $f_{v0}$ without imposing simple assumptions (as spherical or special symmetries) on inhomogeneities. 
First of all, from WMAP data we put a maximum allowed mean value 
for ${\Omega}_{k0}\simeq 0.01$. For ${\Omega}_{m0}$ we consider range of values $(0.2,0.35)$ (see for example \cite{Na}).
In fig \ref{fig1}, \ref{fig2}, \ref{fig3} we plot ${\Omega}_{\Lambda0}$  vs. $f_{v0}$ and ${\Omega}_{m0}$ with ${\Omega}_{k0}=0.01$ and $\{{\epsilon}_i\}=\{0,0.1,0.9\}$.
As apparent from the figures, a set of standard $\Lambda$CDM values for ${\Omega}_{\Lambda0},{\Omega}_{m0},{\Omega}_{k0}$ is compatible with a quite large fraction of voids (roughly up to $0.8$). In particular, we see that the maximum value ${\Omega}_{m0}\simeq 0.35$ together with the maximum allowed mean value 
for ${\Omega}_{k0}$ predicted by WMAP ($\simeq 0.01$) is compatible with the minimum value 
${\Omega}_{\Lambda0}\simeq 0.65$ and a large fraction of voids (up to $\simeq 0.8$). If this were the case, spatial
inhomogeneities could account for a maximum percentage $\approx 10\%$ with respect to the concordance value
${\Omega}_{\Lambda0}\simeq 0.73$.
 Moreover, note that for ${\epsilon}=0 $ or $\sim 0.1$ we have that
huge values for ${\Omega}_{\Lambda 0}>0.8$ can be obtained with ${\Omega}_{m0}\in[0.25,0.35]$ together with
a huge value for $f_{v0}>0.8$. Conversely, for higth $\epsilon\sim 0.9 $ we have that with  ${\Omega}_{m0}\in[0.25,0.35]$,
${\Omega}_{\Lambda0}>\simeq 0.73$ can be obtained for $f_{v0}< \simeq 0.1$.  This can help to understand the interesting role of 
$\epsilon$ in the $\Lambda$FB model. Finally, we consider the case with $\epsilon={\Omega}_{k0}=0$ where a situation 
close to figure \ref{fig1} arises. Note that the concordance model is regained by further setting  
$f_{v0}=0$. As explained at the end of section 2,  this case corresponds to the physical situation where both walls and voids have 
mean zero 
spatial curvature at any time but where voids represent an 
underdensity. Hence, voids expand faster than walls and as a result 
a non trivial volume void fraction $f_v(t)$ with $f_{v0}\neq 0$ emerges as the case with ${\Omega}_{k0}\neq 0$.
\section{Age of the universe compatible with the $\Lambda$FB model}
An important test in cosmology is given by the age of the universe.
The age of globular cluster implies the universe should be certainly older than 
$>12$Gyr. In this section we give an estimation of the age of the universe predicted by $\Lambda$FB model.
First of all, from equation (\ref{13}) evaluated at present time $t_0$ we get:
\begin{equation}
t_0=\frac{2}{3H_0\sqrt{{\Omega}_{\Lambda0}}}\ln\left(
\sqrt{(1-f_{v0})\frac{{\Omega}_{\Lambda0}}{{\Omega}_{m0}(1-{\epsilon}_{i})}}+
\sqrt{(1-f_{v0})\frac{{\Omega}_{\Lambda0}}{{\Omega}_{m0}(1-{\epsilon}_{i})}+1}\right)
\label{time}
\end{equation}
where $H_0$ is the present cosmological constant at the scale of homogeneity.
A consistent value for $t_0$ ($>12$Gyr) is always possible with a large
volume fraction of voids (for some numerical examples see table 2).\\
Let us give an example: let us consider  $t_0\geq 13$ Gyr, a restrictive value for ${\epsilon}_{i} =0$, a large amount of dark energy
(${\Omega}_{\Lambda0}\in[0.67,0.75]$) and a relatively small
curvature, i.e. ${\Omega}_{k0}\leq 0.08$. 
For $H_0$ we consider the range $(55-75)$km/s/Mpc, since a smaller value appears in contrast with the actual data.
From eq.(\ref{time}) we get a reasonable $t_0$ with a void fraction $f_{v0} \approx (40-50)\%$ and $H_0 \approx (55-60)$km/s/Mpc.
Increasing $H_0$ up to $\simeq 70$km/s/Mpc and considering ${\epsilon}_{i}<<1$, we got an age
$t_0 \geq 13$Gyr with  $f_{v0}\approx\in(0,0.3)$.
This shows how the model can account for a reasonable choice of the cosmological parameters (age,$H_0$, curvature, etc..) with 
a fraction of voids as large as $30\%$. 
An even greater fraction of voids, $f_{v0}>0.3$, and $H_0\simeq 70$km/s/Mpc
require a larger value of ${\epsilon}_{i}$ and more precisely ${\epsilon}_{i}\sim f_{v0}$.
\begin{table}[t]
\begin{center}
\begin{tabular}{|c|c|c|c|c|}
\hline
${\Omega}_{\Lambda0}$ & ${\Omega}_{m0}$ & $f_{v0}$ & $H_0$ & $t_0$\\
\hline
$0.67$ & $0.26$ & $0.4$ & $55$ & $15.1$\\
$0.67$ & $0.26$ & $0.4$ & $60$ & $13.9$\\
$0.67$ & $0.26$ & $0.5$ & $55$ & $14.1$\\
$0.67$ & $0.31$ & $0.4$ & $55$ & $14.2$\\
$0.67$ & $0.31$ & $0.4$ & $60$ & $13.0$\\
$0.67$ & $0.31$ & $0.5$ & $55$ & $13.2$\\
$0.67$ & $0.34$ & $0.4$ & $55$ & $13.6$\\
$0.70$ & $0.27$ & $0.4$ & $55$ & $14.9$\\
$0.70$ & $0.27$ & $0.4$ & $60$ & $13.6$\\
$0.70$ & $0.27$ & $0.5$ & $55$ & $13.9$\\
$0.70$ & $0.30$ & $0.4$ & $55$ & $14.3$\\
$0.70$ & $0.30$ & $0.4$ & $60$ & $13.1$\\
$0.70$ & $0.30$ & $0.5$ & $55$ & $13.3$\\
$0.71$ & $0.26$ & $0.4$ & $55$ & $15.0$\\
$0.71$ & $0.26$ & $0.4$ & $60$ & $13.8$\\
$0.71$ & $0.26$ & $0.5$ & $55$ & $14.0$\\
$0.73$ & $0.26$ & $0.4$ & $60$ & $13.7$\\
$0.73$ & $0.26$ & $0.5$ & $55$ & $14.0$\\
$0.75$ & $0.26$ & $0.4$ & $60$ & $13.7$\\
\hline
\end{tabular}
\caption{Age of the universe compatible with $\Lambda$FB model.}
\label{age}
\end{center}
\end{table}
\section{Dressing cosmological parameters and the distance angular function}
One of the issue of the Buchert averaging scheme is how to relate volume average quantities which are non-local
to observable ones, which are local. In contrast to many approaches to the Buchert equations which usually neglect this fundamental issue and following Wiltshire \cite{12,sv}, we match the radial null section of the wall metric
(\ref{1}), rewritten as
\begin{eqnarray}
ds_{w}^2&=&-dt^2+a_{w}^2\left[d{\eta}_w^2+{\eta}_w^2d{\Omega}^2\right]\label{s1}\\
        &=&-dt^2+\frac{{(1-f_v)}^{\frac{2}{3}}a^2}{f_{wi}^{\frac{2}{3}}}
          \left[d{\eta}_w^2+{\eta}_w^2d{\Omega}^2\right].\nonumber
\end{eqnarray}
with the metric at the scale of homogeneity given by
\begin{equation}
ds^2=-dt^2+a^2d{\eta}^2+A(t,\eta)d{\Omega}^2,
\label{s2}
\end{equation}
$A(t,\eta)$ is an area function satisfying
$4\pi\int_0^{{\eta}_{\mathcal{H}}}A d\eta=a^2V_i({\eta}_{\mathcal{H}})$,
where ${\eta}_{\mathcal{H}}$ is the particle horizon radius.
Practically, the wall observer must dress the cosmological parameter and not simply to relate the volume average scale factor to the observed redshift $z$. The dressed wall geometry is:
\begin{equation}
ds_w^2=-dt^2+a^2\left[d{\eta}^2+{(1-f_v)}^{\frac{2}{3}}
f_{wi}^{-\frac{2}{3}}{\eta}_w^2 d{\Omega}^2\right],
\label{s3}
\end{equation}
with
\begin{equation}
d{\eta}_w=\frac{f_{wi}^{\frac{1}{3}}d\eta}{{(1-f_v)}^{\frac{1}{3}}},\;\;
\eta=\int_{t}^{t_0}\frac{dt}{a}.
\label{s4}
\end{equation}
It is by means of the metric (\ref{s3}) that the wall observer, in galaxies,
measures the distance-reashift function $d_L(z)$. The angular-distance relation
$d_A(t)$
(remember that $d_L=(1+z)^2 d_A$) is:
\begin{eqnarray}
& &d_A(z)=\frac{a_0}{(1+z)}\overline{\eta}_w,\;\;\;1+z=\frac{a_0}{a}\label{s5},\\
& &{\overline{\eta}}_w={(1-f_v)}^{\frac{1}{3}}
\int_t^{t_0}\frac{dt}{{(1-f_v)}^{\frac{1}{3}}a}.\nonumber
\end{eqnarray}
Using eq.(\ref{13}), expression (\ref{s5}) becomes:
\begin{equation}
d_A(t)={\sinh}^{\frac{2}{3}}\left(\frac{3}{2}H_0\sqrt{{\Omega}_{\Lambda0}}t\right)
\int_t^{t_{0}}\frac{dt}{{\sinh}^{\frac{2}{3}}\left(\frac{3}{2}H_0\sqrt{{\Omega}_{\Lambda0}}t\right)},
\label{s6}
\end{equation}
where $t_0$ is given by eq.(\ref{time}).
We can see that $d_A$ has the same functional form versus the time $t$ both for
$\Lambda$CDM and $\Lambda$FB models. The only difference is the expression of the function $t_0$.
Obviously the expression for $d_A$ changes with respect to the concordance model when 
expressed in terms of the redshift $z$. To express $t(z)$ along the past null 
cone, we must to know the function
$a(t)$ which, for the $\Lambda$FB model can be obtained by integrating numerically the equations
(\ref{9}). 
\section{Conclusions}
We have presented a preliminary study of the recent $\Lambda$FB cosmological model.
An explicit formula relating all the cosmological parameters of the $\Lambda$FB model is obtained
and analyzed. The relevant feature of the $\Lambda$FB model is the presence of the cosmological constant and 
spatial inhomogeneities without spherical symmetry \cite{aer}. 
By consequence, the standard
$\Lambda$CDM model can be recovered with a suitable choice of the parameters.
The model allows to analyse the departures from the standard cosmological model without invoking
perturbation theory.\\
The first result of the present study is  the consistency of a large 
volume voids fraction ($>0.1$) with a small spatial curvature, even  within WMAP constraint
($|{\Omega}_{k0}|\leq 0.01$). This falsifies the argument
often used in the literature to rule out dark energy, i.e. a large fraction of voids observed 
necessarily implies a large negative value for the curvature .\\
The second result is the important role (absent in the FB model) of the initial volume void fraction $f_{vi}$. 
In fact, for a reasonable range of values for the cosmological parameters
${\Omega}_{m0},{\Omega}_{k0},{\Omega}_{\Lambda0}$ and dependent on the
parameter ${\epsilon}_i$, the model generally provides two possibles values for the current volume void fraction: 
$f_{v01}<<1$ and $f_{v02}\geq 0.1$. If the initial value
$f_{vi}<f_{v01}$, then the universe evolves up to a final volume void fraction $f_{v01}$, while for
$f_{vi}>f_{v01}$ the system evolves up to the second root $f_{v02}$. 
As it is evident from the table
(\ref{voids}), the model is quite sensitive to small variations of cosmological parameters.
Finally, we analysed the formula (\ref{L1}) giving the exact value
for ${\Omega}_{\Lambda0}$ in terms of the other current cosmological parameters. Setting the
maximum value allowed for ${\Omega}_{k0}\simeq 0.01$ and ${\Omega}_{m0}\simeq 0.35$, we get 
(see figures \ref{fig1}, \ref{fig2}, \ref{fig3})
that the lower value for
${\Omega}_{\Lambda0}\simeq 0.65$ is compatible with a volume void fraction $\in (0,0.8)$. 
Hence, considering a total amount of dark energy $73\%$ predicted by the $\Lambda$CDM model, 
the inhomogeneities could account for a maximum percentage $\approx 10\%$ of ${\Omega}_{\Lambda0}$. \\
As a final consideration, note that also in the $\Lambda$FB model it is possible to add clock effects present in 
\cite{12}. This can be done simply by considering $I_w$ as a lapse function together with $\epsilon<<1\sim f_{vi}$ 
(see \cite{sv}). Equations (\ref{8}), (\ref{9}) and formula (\ref{15}) are left unchanged. The  changes are in the dressing procedure, i.e.
equations (\ref{s1})-(\ref{s6}). Moreover, in the case of non vanishing lapse function the cosmological parameters must be 'dressed' 
\cite{10,12}. As an example, if clock effects are present, ${\Omega}_{m}$  is the bare volume-average density parameter while the measured 
density parameter in walls ${\Omega}_{mw}$ is given by ${\Omega}_{mw}=I_w^3{\Omega}_m$.
This is a preliminary parametric study on the relation between cosmological parameters at the present time. 
In a future paper we intend to study the complex problem to fit the observational data by integrating numerically the model field equations.

\section*{Acknowledgements}
We would like to thank D. L. Wiltshire  for useful comments and suggestions.


\begin{thebibliography}{0}
\bibitem{1}B.S. Perlmutter, et al., {\it Astrophys. J}. {\bf 483} (1997) 565. 
\bibitem{q}A.G. Riess, et al., {\it Astron. J.} {\bf 116}  (1998) 1009.
\bibitem{2}B.S. Perlmutter, et al., 
{\it Astrophys. J.} {\bf 517} (1999) 565.
\bibitem{16}M.N. C{\ac{e}}l{\ac{e}}rier,   
{Astron. and Astrophys.} {\bf 353} (2000) 63.
\bibitem{To}M. Tanimoto and Y. Nambu, {\it Class. Quantum Grav.} {\bf 24} (2007) 3843.
\bibitem{13}J.W. Moffat,   
{\it JCAP} {\bf 0605} (2006) 001. 
\bibitem{123}K. Enqvist,  
{\it Gen. Relativ. Gravit.} {\bf 40} (2008) 451.
\bibitem{13BB}H. Alnes, M. Amarzguioui, and O. Gron,   
{\it Phys. Rev. D}{\bf 73} (2006) 083519.
\bibitem{z1}W. Valkenburg, {\it JCAP}{\bf 0906}(2009) 010.
\bibitem{z2}T. Biswas, A. Notari and  W. Valkenburg,  {\it JCAP}{\bf 1011}(2010) 030
\bibitem{z3}W. Valkenburg, {\it JCAP}{\bf 1201}(2012) 047.
\bibitem{20BB}K. Tomita, {\it Prog. Theor. Phys.} {\bf 106} (2001) 929.
\bibitem{3}A. Paranjape, and T.P. Singh,  {\it Class. Quantum Grav.} {\bf 23} (2006) 6955.
\bibitem{3B}V. Marra, E.W. Kolb, S. Matarrese, and A. Riotto,   
{\it Phys. Rev. D} {\bf 76} (2007) 123004.
\bibitem{31bis}T. Buchert, and M. Carfora, 
{\it Class. Quantum Grav.} {\bf 19} (2008) 195001.
\bibitem{3BB}V. Marra, E.W. Kolb, and S. Matarrese,  
{\it Phys. Rev. D}{\bf 77} (2008) 023003.
\bibitem{4}S. R\"as\"anen,   
{\it JCAP}{\bf 0611} (2006) 003.
\bibitem{5}T. Buchert, 
{\it Gen. Relativ. Gravit.} {\bf 32} (2000) 105.
\bibitem{500}T. Buchert,  
{\it Gen. Relativ. Gravit.} {\bf 40} (2008) 467.
\bibitem{7}T. Buchert, and M. Carfora,   
{\it Phys. Rev. Lett.} {\bf 90} (2003) 031101.
\bibitem{8}T. Buchert, and M. Carfora,    
{\it Class. Quantum Grav.} {\bf 19} (2002) 6109.
\bibitem{10}D.L. Wiltshire,   
{\it Phys. Rev. Lett.} {\bf 99} (2007) 251101.
\bibitem{12}D.L. Wiltshire,   
{\it New J. Phys.} {\bf 9} (2007) 377.
\bibitem{ww1} J. Kwan, M. J. Francis and  G. F. Lewis, {\it Mon. Not. R. Astron. Soc. Lett.}{\bf 399} (2009)L6.
\bibitem{ww2} P.R. Smale and D.L. Wiltshire, {\it  Mon. Not. R. Astron. Soc.}{\bf 413}  (2011) 367.   
\bibitem{z1}R.A. Vanderveld, \'E.\'E. Flanagan, and I. Wasserman, 
{\it Phys. Rev. D}{\bf 74} (2006) 023506.
\bibitem{E3}A.E. Romano, {\it JCAP}{\bf 05} (2010) 020.
\bibitem{cielo}A. Krasinski, C.Hellaby, K. Bolejko, and M.N. C{\ac{e}}l{\ac{e}}rier, 
{\it General Relativ. Gravit.} {\bf 42} (2010) 2453.
\bibitem{C2}M.N. C{\ac{e}}l{\ac{e}}rier, K. Bolejko, and A. Krasinski,  
{\it Astron. Astrophys} {\bf A21} (2010) 518.
\bibitem{web}P.M.Sutter, et al., Astrophys. J. {\bf 761}, 44 (2012).
\bibitem{100}F. Hoyle and M.S. Vogeley, {\it Ap. J.} {\bf 607} (2004) 751.
\bibitem{100v}D.C. Pan, M.S. Vogeley,  F. Holey, Y-y Choi and C. Park, arXiv:1103.4156.
\bibitem{101v}A.V. Tikhonov and I.D. Karachentsev, {\it Astrophys. J.} {\bf 653} (2006) 369.
\bibitem{102}M. Zaldarriaga, D.N. Spergel and U. Seljak, {\it Astrophys. J.} {\bf 488} (1997) 1.
\bibitem{103}E. Komatsu et al., {\it Astrophys. J.S.} {\bf 180} (2009) 330.
\bibitem{104}J.M.Virey at al.., {\it JCAP} {\bf 12} (2008) 008. 
\bibitem{u1}L. Amendola, K. Kainulainen, V. Marra and V. Quartin, {\it Phys. Rev. Lett.}
{\bf 105} (2010) 121302.
\bibitem{u2}V. Marra and M. Paakkonen, {\it JCAP}{\bf 12} (2010) 021.
\bibitem{aer}A.E. Romano and C. Pisin{\it JCAP} {\bf 016}, (2011) 1110. 
\bibitem{void}V. Marra and A. Notari, {\it Class. Quant. Grav.} {\bf 28} (2011)164004.
\bibitem{sv}S. Viaggiu, {\it Class. Quantum Grav.} {\bf 29} (2012) 035016.
\bibitem{sv2}S. Viaggiu, {\it Class. Quantum Grav.} {\bf 27} (2010) 155002.
\bibitem{106}J.N. Dossett and M. Ishak, arXiv:1205.2422.
\bibitem{107}A.H Guth and Y. Nomura, arXiv:1203.6876. 
\bibitem{190}Y. Wang and P. Mukherjee, {\it Phys. Rev. D} {\bf 76} (2007) 103533.
\bibitem{106T}T. Clifton, P.G. Ferreira and J. Zuntz, {\it JCAP} {\bf 07} (2009) 029. 
\bibitem{Na}J. A. Peacock et al., {\it Nature} {\bf 410} (2001) 169.
\end{thebibliography}
\end{document}